# Density-functional study of $Li_xMoS_2$ intercalates ($0 \leq x \leq 1$)

*Andrey N. Enyashin,*[a,b], and Gotthard Seifert[a]*

[a] Physical Chemistry Department, Technical University Dresden, Bergstr. 66b, 01062 Dresden, Germany

[b] Institute of Solid State Chemistry UB RAS, Pervomayskaya Str. 91, 620990 Ekaterinburg, Russia

Enyashin@ihim.uran.ru; Gotthard.Seifert@chemie.tu-dresden.de

The stability of Lithium intercalated 2H- and 1T allotropes of Molybdenum disulfide ($Li_xMoS_2$ ) is studied within a density-functional theory framework as function of the Li content (x) and the intercalation sites. Octahedral coordination of Li interstitials in the van der Waals gap is found as the most favorite for both allotropes. The critical content of Lithium, required for the initialization of a 2H→1T phase transition is estimated to $x \approx 0.4$. For smaller Li contents the hexagonal 2H crystal structure is not changed, while 1T-$Li_xMoS_2$ compounds adopt a monoclinic lattice. All allotropic forms of $Li_xMoS_2$ - excluding the monoclinic $Li_{1.0}MoS_2$ structure - show metallic-like character. The monoclinic $Li_{1.0}MoS_2$ is a semiconductor with a band gap of 1.1 eV. Finally, the influence of Li intercalation on the stability of multiwalled $MoS_2$ nanotubes is discussed within a phenomenological model.

*Corresponding author:

e-mail: Enyashin@ihim.uran.ru

Phone: +7 (343) 362 3115

Fax: +7 (343) 374 4495

Address: Institute of Solid State Chemistry UB RAS, Pervomayskaya Str. 91, 620990 Ekaterinburg, Russia

## 1. Introduction

Layered transition-metal dichalcogenides have been the subject of numerous studies as materials with distinct tribological, photoelectrical, optical and catalytical properties.[1-4] The layer structure allows also the intercalation of atoms, molecules or ions between the layers.[5] Of special interest is the intercalation with Lithium, which can open the application as electrode materials for batteries with a high energy storage. Although, the intercalation chemistry of layered chalcogenides has been studied experimentally since the 1970's, a profound theory of the structure and the properties of their Lithium intercalates is lacking. The rich polytypism and phase transitions as a consequence of the charge transfer may hinder to some extend the application of the layered chalcogenides as electrode materials.[6,7] The fundamental understanding of these phenomena may improve the value of these materials for electrochemical utilizations.

Molybdenum disulfide $MoS_2$ is a good representative compound of the family of layered chalcogenides.[1] Like in the structure of other layered dichalcogenides the metal atoms in $MoS_2$ have a six-fold coordination environment. Depending on the arrangement of the S atoms two kinds of the hexagonal S-Mo-S triple layers are possible, which are composed by either prismatic $D_{3h}$- or octahedral $O_h$-$MoS_6$ units. Such triple layers interact by weak van-der-Waals interactions and may be stacked in different ways. For example, natural $MoS_2$ occurs as a mixture of two stable polymorphs based on $D_{3h}$-$MoS_6$ units: the hexagonal 2H-$MoS_2$ and the rhombohedral modification 3R-$MoS_2$, which unit cells include two or three layers, respectively.[8] The 2H-polytype is more stable, hence the 3R-polytype transforms into the 2H structure upon heating.

The $MoS_2$ polytype based on $O_h$-$MoS_6$ units has not yet been found in the nature and is viewed as an unstable modification. Nevertheless, S-Mo-S layers with an octahedral coordination of metal atoms can be synthesized by intercalation of 2H-$MoS_2$ by alkali-metals Li and K, forming 1T-phases of $Li_xMoS_2$ and 1T-$K_xMoS_2$ in analogy to the stable allotropes of hexagonal 1T-dichalcogenides (e.g. $TiS_2$).[9-11] The origin of this phase transition has been explained for the pure bulk hexagonal 2H- and 1T-$MoS_2$ allotropes in the framework of both the ligand field and band theory using semi-empirical[12] and *ab initio* methods.[13,14] However, subsequent solvation and reduction of 1T-$MoS_2$ intercalates leads to the exfoliation and release of free 1T-$MoS_2$ layers,[15-17] which depending on the composition of the intercalate and reducing agent were found to be restacked in various superstructures as revealed by Raman-scattering,[18] electron and x-ray diffraction[18-20] and scanning tunneling microscopy.[21] It suggests that the phase diagram of hexagonal dichalcogenides with partially occupied *d*-shells may be more complex and may contain monoclinic phases with a considerably distorted octahedral environment of metal atoms.

Molybdenum disulfide upon Lithium intercalation is an example for the study of the destabilization of a dichalcogenide lattice. Currently, experimental data combined with the density-functional theory calculations have been applied for the study of stoichiometric $LiMoS_2$ intercalates with different intra- and interlayer stacking patterns of Mo atoms with octahedral coordination.[22-24] These calculations evidenced a high stability of monoclinic $LiMoS_2$ structures in comparison to the ideal hexagonal 1T-$LiMoS_2$ structure.

To our knowledge, there exist only very few theoretical studies of $Li_xMoS_2$ intercalates with varying Li content for the hexagonal 2H-$MoS_2$ host lattice. The evolution of the "charge capacity" as a function of the composition was studied by means of a cluster model within a semi-empirical approach.[25] Density-functional theory calculations were performed to estimate the volume expansion and the intercalation energy for a set of 2H-$Li_xMoS_2$ compositions.[26]

In the present work the stability, the electronic structure and the crystal geometry for 2H-$Li_xMoS_2$ and 1T-$Li_xMoS_2$ intercalates are investigated in the range of the Li content $0 \le x \le 1$ and for different intercalation sites, using a DFT approach. Additionally, a phenomenological model implying available experimental and calculated data is developed to estimate the possible influence of Lithium intercalation on the stability of multiwalled $MoS_2$ nanotubes.

## 2. Computational details

The calculations are performed using the SIESTA 2.0 implementation[27,28] within the framework of the density-functional theory (DFT).[29] Initial test calculations have been performed for the bulk structures of 2H-$MoS_2$ and bcc-Lithium, using the exchange–correlation potentials within the local-density approximation (LDA) with the Perdew-Zunger parametrization[30] and the generalized gradient approximation (GGA) using parameterization proposed by Perdew–Burke–Ernzerhof.[31] However, these tests gave very similar results for LDA and GGA in the calculations of the band structure and the lattice parameters as well as for the estimations of the relative stability of 1T and 2H-$MoS_2$ allotropes. Therefore, only the exchange–correlation potentials treated within LDA are applied for the calculations of all $MoS_2$ and $Li_xMoS_2$ structures. The core electrons are treated within the frozen core approximation, applying norm-conserving Troullier–Martins pseudopotentials.[32] The valence electrons were taken to be $3s^23p^4$ for S, $5s^14d^5$ for Mo and $2s^1$ for Li. Initially, the pseudopotential core radii were chosen, as suggested by Martins, and to 1.70 $a_B$ for S and 2.45 $a_B$ for *s*-, *d*- and 2.65 $a_B$ for *p*-states of Mo, and 2.45 $a_B$ for the *s*-states of Li. Basis sets - single-ζ, double-ζ with and without polarization functions - were tested on the bulk 2H-$MoS_2$ and bcc-Lithium. These test calculations demonstrated a good suitability of the initial set of core radii for the $MoS_2$ system for different types of basis sets. However, a simultaneous description of both, the band structure and the lattice parameters was not possible for bcc-Lithium at any basis set. In order to reproduce as accurately as possible the experimental lattice parameters and band structure of bcc-Lithium the core radii for the generation of the Li pseudopotential were varied and found to be optimal at a value of 3.50 $a_B$. Furthermore, in all calculations of $Li_xMoS_2$ structures only a double-ζ basis set is used for all atoms together with "standard" pseudopotentials for Mo and S and the modified pseudopotential for Li, as described above.

For *k*-point sampling, a cutoff of 10 Å was used,[33] which gave 58 independent *k* points in the first Brillouin zone for the unit cell of 2H-$MoS_2$. The *k*-point mesh was generated by the method of Monkhorst and Pack.[34] The real-space grid used for the numeric integrations was set to correspond to an energy cutoff of 200 Ry. All calculations were performed using variable-cell and atomic position relaxation, with convergence criteria set to correspond to a maximum residual stress of 0.1 GPa for each component of the stress tensor, and maximum residual force component of 0.01 eV/Å.

Host lattices of $Li_xMoS_2$ intercalates are considered using two main atomic models (Fig. 1). First, 1×1×1 and 1×1×2 cells for 2H- and 1T-allotropes with composition $Mo_2S_4$ are selected, respectively. It allows to study the hexagonal phases with the Li content $x$ = 0.0, 0.5 and 1.0. Second, the hexagonal supercells 2×2×1 and 2×2×2 for 2H- and 1T-allotropes with composition $Mo_8S_{16}$ are chosen for the study of $Li_xMoS_2$ intercalates in a wide range of Li content and with different occupation of intercalation sites, which allow as well to trace a possible transformation of the lattice symmetry.

## 3. Results and Discussion

### 3.1. Properties of $MoS_2$ allotropes

As a first step we have performed the calculations of the lattice parameters and band structure for the pure intercalation matrices. The parent 2H- and 1T-$MoS_2$ allotropes, as well as monoclinic $MoS_2$ were considered. The lattice parameters *a* and *c* obtained at the equilibrium volume in the present work for the 2H-$MoS_2$ allotrope are equal 3.12 Å and 11.74 Å, which is in good agreement with experimental data and are similar to those reported in previous theoretical works.[35,36] The geometry of the ideal 1T-$MoS_2$ allotrope with hexagonal lattice has not been reported before, and our calculated geometry demonstrate a slight difference in the interatomic distances comparing with 2H-$MoS_2$. The calculated lattice parameters are $a$ = 3.14 Å and $c$ = 5.62 Å. For the further discussion of the possible influence of the polytypism on the Lithium intercalation sites, we have performed also calculations for hypothetical hexagonal 2T-$MoS_2$ polytype, which contains two antiparallel layers within the unit cell. The optimized parameters are: $a$ = 3.11 Å and $c$ = 11.47 Å. I.e., it has similar intralayer bond lengths and interlayer distance as in 1T-$MoS_2$. In spite of relatively close lattice parameters of 2H- and 1T-$MoS_2$, the change of the coordination polyhedra from trigonal prisms $D_{3h}$-$MoS_6$ to octahedra $O_h$-$MoS_6$ crucially changes the electronic properties and the relative stabilities of these allotropes.

In agreement with previous theoretical and experimental data[5] both 1T- and 2T-$MoS_2$ allotropes with an octahedral coordination of the Mo atoms have been found to be considerably less stable, than the 2H-$MoS_2$ allotrope by 0.80 eV/$MoS_2$ (for comparison, 0.78 eV/$MoS_2$ in GGA approach). Many aspects in the stability of pure and Li-intercalated phases are determined by their electronic structure. The calculated electronic band structures for the hexagonal 1T- and 2H-$MoS_2$ allotropes are drawn on Fig. 2.1. The 2H-$MoS_2$ allotrope is a semiconductor with a calculated direct band gap 2.18 eV and indirect $\Gamma \rightarrow ½K\Gamma$ gap 0.57 eV. In the calculated electronic structure of this phase three bands can be clearly distinguished, which agrees with the results of former experimental and theoretical investigations.[12,37] The valence band composed mainly S*3p*-states, about ~3.5-6.0 eV below the Fermi lenergy. The states just below the Fermi level are Mo*4d*-states. The bottom of the conduction band is also dominated by Mo*4d*-states. In terms of ligand field theory the semiconducting nature of 2H-$MoS_2$ is caused by the splitting of the Mo*4d*-orbitals of a $D_{3h}$-$MoS_6$ unit into the occupied Mo$4d_{z^2}$ states and the unoccupied Mo$4d_{xy}$ , Mo$4d_{x^2-y^2}$ , Mo$4d_{xz}$ and Mo$4d_{yz}$ states.

In turn, the 1T-$MoS_2$ allotrope shows a metal-like character (see Fig. 2.2) as discussed already in the literature.[13,14] Like in the case of 2H-$MoS_2$ the valence band of the S*3p*-states is about ~3 eV below the Fermi level. Though, the Mo*4d*-states in 1T-$MoS_2$ form a band, which hosts the Fermi level. In terms of crystal field theory the origin of this band can be described by the splitting of the Mo*4d*-orbitals in an octahedral ligand field into three degenerated Mo$4d_{xy,yz,xz}$-orbitals, occupied with only two electrons, and the unoccupied Mo$4d_{z^2}$ and Mo$4d_{x^2-y^2}$ levels.

The incomplete occupation of Mo$4d_{xy,yz,xz}$-orbitals by two electrons in 1T-$MoS_2$ leads to the metallic ground state. Therefore, the 1T-$MoS_2$ lattice can be stabilized by addition of electrons, e.g. from Li atoms. On the contrary, additional electrons in the semiconducting 2H-$MoS_2$ allotrope have to occupy the Mo$4d_{xy}$ and Mo$4d_{x^2-y^2}$ orbitals, which should result in a metallic-like character and cause destabilization of the lattice.

However, the layers exfoliated from alkali-metal intercalated $MoS_2$ phases show a non-hexagonal (distorted) 1T-$MoS_2$ lattice structure with various regular patterns of Mo atoms.[19-21] Thus, additionally, we have studied also monoclinic $MoS_2$ phases. The atomic positions of the Mo atoms in the 2×2×1 supercell of the hexagonal 1T-$MoS_2$ have been changed along initially equal *a* and *b* lattice vectors in such way, that a rhombus of Mo atoms – $Mo_4$ – became clearly discernible within the $MoS_2$ layer. Two polytypic layered structures containing two $MoS_2$ layers have been considered. In first structure the $Mo_4$ rhombi of two layers were placed above each other, while in the second structure the atomic coordinates of two layers were displaced relative each other on *a*/2 and *b*/2. The optimization of these structures has evidenced the existence of at least two monoclinic phases with slightly different formation energies and lattice parameters: (i) monoclinic $MoS_2$ phase with lattice parameters $a$ = 6.33 Å, $b$ = 6.51 Å and $c$ = 10.95 Å, ~1.12 eV/$MoS_2$ less stable than hexagonal 2H-$MoS_2$, and (ii) monoclinic polytype with lattice parameters $a$ = 6.35 Å, $b$ = 6.52 Å and $c$ = 10.78 Å,

~1.02 eV/$MoS_2$ less stable than hexagonal 2H-$MoS_2$. Both polytypes show the relaxation of the Mo sublattice within the layers into $a \times a\sqrt{3}$ superstructure, composed of zigzag-like chains with Mo-Mo distances 2.76 – 2.78 Å, which is much shorter than the ~3.1 Å in hexagonal 2H- or 1T-$MoS_2$.

Thus, both monoclinic $MoS_2$ phases are found even less stable than the ideal hexagonal 1T-$MoS_2$, which is ~0.2-0.3 eV/$MoS_2$ less stable than hexagonal 2H-$MoS_2$. Possible way for stabilization of a distorted 1T-$MoS_2$ can be understood as well from the electronic structure. The band structure of the more stable monoclinic $MoS_2$ polytype is shown in Fig. 2.3. The metallic properties of this phase with octahedral coordination of Mo atoms are maintained as for 1T-$MoS_2$. However, in the region between 1-1.5 eV above Fermi level, a splitting of the Mo*4d*-bands is visible, which can be attributed to the removal of the degeneracy of Mo$4d_{xy,yz,xz}$-orbitals. The resulting band is splitted at many k-points. Additional electrons, e.g. from an intercalated atom, could transform the band structure in such way that complete opening of a gap would be possible and a semiconducting intercalate could exist at a certain amount of an intercalant. It may lead to the change in the relative stability of hexagonal and monoclinic 1T-$MoS_2$ lattices between pure and intercalated compounds. This statement is supported by the results of the calculations of Lithium intercalated $MoS_2$ compounds, discussed below.

### 3.2. Properties of hexagonal $Li_{0.5}MoS_2$ and $Li_{1.0}MoS_2$ intercalates

All allotropes of $MoS_2$ possess octahedral and tetrahedral interstitials between sulfur atoms at the van der Waals gap. Due to sterical factors the Li accommodation could be more favoured in an octahedral sulfur environment. However, Lithiumsulfide - $Li_2S$ - has an anti-fluorite structure, composed of $LiS_4$ tetrahedra.[38] Thus, a first step of the study of the Li intercalation into an $MoS_2$ host lattice should be the determination of the most favourable interstitial site within the van der Waals gap. It can prove, whether the intercalation process of a $MoS_2$ hosting matrix is more dependent on the sterical factor or electronic factors play a more important role in the bonding interaction between a guest atom and hosting matrix.

These preliminary calculations have been performed for hexagonal unit cells of 2H-, 2T- and 1×1×2 1T- $Li_xMoS_2$ intercalates with x = 0.5 and 1.0. The stability of an intercalate was characterized using the energy of formation from the bulk structures of molybdenum disulfide 2H-

$MoS_2$ and metallic Lithium with bcc-structure[39] according to the reaction: 2H-$MoS_2$ + $x$ Li → $Li_xMoS_2$.

The geometries of the optimized unit cells for all these hexagonal $Li_xMoS_2$ intercalates show, that intercalation does not change crucially the intralayer bond lengths (Table 1). The lattice parameters $a$ of all intercalates are slightly increased, compared to those of the pure 2H- and 1T-$MoS_2$ phases. The difference does not exceed 0.04 Å and 0.06 Å, respectively. The interlayer $MoS_2$ distances are essentially increased by 1-1.2 Å for all cases. All main changes in the relative stability of these allotropes upon Lithium intercalation may be explained again mainly by the electronic structure.

Indeed, while the formation energies for all 2H-$Li_xMoS_2$ intercalates have values of ~-0.7…-0.8 eV/Li-atom, independent from the composition, the formation energies of 1T-$Li_xMoS_2$ intercalates show a considerable stabilization of the 1T-$MoS_2$ lattice with the increase of the Li content (Table 1). For a maximum Li content $x$ = 1.0, the relative stabilities of the 2H- and 1T-$MoS_2$ allotropes exchange, and the 1T-$Li_{1.0}MoS_2$ gains already ~0.1 eV/Li-atom over 2H-$Li_{1.0}MoS_2$ and has the formation energy ~ -0.9 V/Li-atom. This phenomenon is consistent with the picture of the electronic structure for these allotropes as discussed and can be described within the framework of the rigid band model (Fig. 2). Consecutive population of the conduction band of Mo*4d* states in 2H-$Li_xMoS_2$ allotropes does not promote any stabilization of the host lattice as it does for 1T-$Li_xMoS_2$ allotropes, where three partially occupied Mo$4d_{xy,yz,xz}$ states near the Fermi level readily accept one more electron.

Although, the comparison of the formation energies for intercalates of the same allotropic modification demonstrates that the sterical factor plays a dominant role for the occupation of interstitial sites by Lithium (Table 1). For all $MoS_2$ host lattices the octahedral cavity in van der Waald gap is found to be the most favorable position for the Li atom. For the 2H-$Li_xMoS_2$ polytypes, where the electron donation from Li to Mo*4d*-levels is suppressed, the energetic differences between the octahedral and tetrahedral coordination of the intercalated Li atoms are less pronounced and do not exceed 0.1 eV/Li-atom. For 1T- and 2T-$Li_xMoS_2$ intercalates these values can differ on ~0.5 eV/Li-atom, indicating that octahedral coordination of Li atoms in this case is favored and assists in the charge transfer.

### 3.3. Properties of $Li_xMoS_2$ intercalates ($0.0 \leq x \leq 1.0$)

The consideration of low Li concentrations requires the consideration of larger cells. Supercells with dimensions 2×2×1 and 2×2×2 for 2H- and 1T-$Li_xMoS_2$ allotropes have been used, which contain up to 8 Li atoms (Fig. 1). Since many variants for the arrangement of the Li atoms at the interstitial sites of the van der Waals gap are possible in both allotropes we have focused our attention only on the octahedral sites, as the most favorable intercalation sites. For these sites we have considered all possible distributions of Li atoms for a given Li content, that can be modeled for the supercells chosen. In total, 22 model systems have been studied for every allotrope (see Supporting Information).

The geometry optimization of these supercells allowed also trace possible changes in the lattice symmetry of the $Li_xMoS_2$ compounds. For intercalates based on 2H-$MoS_2$ no remarkable deviation from the hexagonal symmetry of the supercell was found almost for the full range of Li content. However, at the maximum intercalation rate ($x$ = 1.0) the host lattice of 2H-$MoS_2$ was transformed into an orthorombic cell with the parameters $a$ = 3.22 Å, $b$ = 5.62 Å and $c$ = 13.27 Å (see Fig. 3.1). The $MoS_6$ prismatic polyhedra composing the layers in this structure are distorted, and a clear alternation in the valence angles and the interatomic distances can be seen. The Mo atoms within a layer of this structure are arranged into zigzag-like chains with a Mo-Mo distance of 2.89 Å, which is shorter than 3.12 Å in the pristine 2H-$MoS_2$. Already these structural transformations indicate a low stability of 2H-$MoS_2$ with a high Li content and reflect a tendency for a chemical disintegration of this phase. One can notice a visual similarity between such distorted 2H-$MoS_2$ layer and the structure of the sulfur-rich surface (001) presented in the molybdenum sesquisulfide ($Mo_2S_3$), where Mo has only oxidation state of +3.[40]

In the case of Li intercalates, based on the hexagonal 1T-$MoS_2$ lattice, a considerable distortion of the structure can be seen already at the smallest considered Li content ($x$ = 0.125). The optimized structures for all 1T-based $Li_xMoS_2$ intercalates can be described within a monoclinic symmetry (see Fig. 3.2 and Supporting Information). Though, these $Li_xMoS_2$ compounds possess another type of superstructural ordering ($2a$×$2a$) differing from the Li-free monoclinic $MoS_2$ structures (see above). They contain "diamond-like" cluster chains of Mo atoms. For example, the layers of $Li_{1.0}MoS_2$ contain the chains of $Mo_4$ rhombi with an Mo-Mo side length of 2.90 Å connected via an Mo-Mo bond with the length of 3.00 Å. This finding fits with the results of previous theoretical and experimental investigations on $LiMoS_2$ compounds, where the same type of superstructure was found as the most stable one.[22]

Thus, the calculations show, that there can be the competitive existence or coexistence of four allotropic forms of $Li_xMoS_2$ intercalates with different structures of the host $MoS_2$ lattice. The stability of all these allotropes can be characterized by their formation energies. The calculated formation energies are shown for the energetically most favored Li arrangements in Fig. 4. (The complete set of energies is given in the Supporting Information). We find - depending on the Li content - two most stable modifications of $MoS_2$ matrices for $Li_xMoS_2$ intercalates should occur. At $x < 0.4$ the undistorted 2H-$MoS_2$ hosting lattice remains as the most favorable one, whereas at higher intercalation rates the monoclinic lattice is stabilized. Of course, our estimations are valid only in the thermodynamic equilibrium for a regular distribution of Li atoms within the $MoS_2$ matrices. Most probably, under experimental conditions, Li intercalation of $MoS_2$ would lead to the appearance of a multiphase sample with variable and diffusion controlled distribution of the Li atoms. However, the calculated concentration limit of Li for the 2H↔1T phase transition correlates well with the available experimental data. XPS measurements on $MoS_2$ single crystals annealed after Li deposition give an evidence that this limit is at $x > 0.2$.[41] Differences in the Raman spectra of $Li_xMoS_2$ has been observed for intercalation rates between $x = 0.1$ and $x = 0.3$.[42] Electrochemical measurements of the incremental capacity have demonstrated, that $MoS_2$ cathode material exhibits a complex intermediate behavior. At least four states can be observed up to the composition $Li_{1.0}MoS_2$, and one of them appears at $x \approx 0.25$.[43]

Noteworthy, the stability curves shown in Fig. 4 for the $Li_xMoS_2$ intercalates with different structural types have different slopes. Obviously, at small intercalation rates both processes – intercalation and deintercalation – can be easily performed without an essential energy expense per Li-atom, since the formation energies for 2H-$Li_xMoS_2$ compounds in this region are nearly independent on x. On the contrary, the energy for monoclinic $Li_xMoS_2$ phases dominating at $x > 0.4$ has a negative slope. These compounds can be characterized by an exothermic process of intercalation, while their deintercalation would be endothermic and requires about 0.8 eV/Li-atom to reversibly achieve the composition with $x < 0.4$.

The increased stability of the monoclinic phases with higher Li content can be understood using the electronic structure of the $Li_{1.0}MoS_2$ intercalate, which is a semiconductor with a band gap of 1.1 eV (Fig. 2.6), in agreement with previous calculations for this particular composition.[22,23] Any Li deintercalation of this compound would lead to the shift of the Fermi level into the valence band and to the destabilization of the electronic system. The most prominent proof of that was given above for the examples of Li-free $MoS_2$ monoclinic allotropes, which are metallic and found to be even less stable than 1T-$MoS_2$. Noticeably, the electronic structure of the 1T-based $Li_{1.0}MoS_2$ intercalate is very similar to that of monoclinic $ReS_2$ or $ReSe_2$, which are genuine semiconductors with the same monoclinic structure of the layers containing "diamond-like" chains of $Re_4$ clusters, since these clusters contain the same number of valence electrons.[12]

While the structure of the hexagonal lattices of 2H- and 1T-$Li_xMoS_2$ intercalates can be considerably distorted with the formation of Mo-Mo bonds, the changes in the volumes of their unit cells, depending on the composition are rather smooth functions with similar slopes (Fig. 5). The main contribution in the volume increase gives the increase of interlayer distance upon Li intercalation.

### 3.4. Stability of Li intercalated $MoS_2$ nanotubes

Recently, nanostructuring has been considered as one of the promising routes to increase the lithium storage capacity of layered chalcogenides as electrode material. While exfoliated $MoS_2$ and $WS_2$ nanoplates - consisting of disordered graphene-like layers and nanoflakes - show a reasonable high and reversible Li storage capacity,[44-46] obviously, an even larger lithium insertion may be achieved using chalcogenide nanostructures with cavities. A few attempts have been performed to fabricate Lithium and other alkaline metal intercalated $WS_2$ and $TiS_2$ nanotubes [47,48] and corresponding fullerene-like nanoparticles.[49,50] Already preliminary studies of the electrochemical behavior of electrodes made of $MoS_2$ and $WS_2$ nanotubes have demonstrated their high lithium storage potential, corresponding to a few mol of Li per mole of sulfide nanotubes.[48] Such high capacity can be attributed to Lithium diffusion into the cavities of nanotubes as well as the intercalation into van der Waals gap between the sulfide layers like in the bulk material. The latter may increase the strain energy of the nanotubes due to the found expansion of the interlayer distance and lead to their exfoliation. At present, this effect is not studied for the sulfide nanotubes. The only related X-ray and TEM studies have been performed on allied $MoS_2$ and $WS_2$ fullerene-like particles. They show an expansion of the interlayer spacing as a function of the radius of inserted alkaline metal atoms, which may be viewed as an indication for the interlayer intercalation.[50] Although, the morphology of these fullerene-like particles prevents an intercalation and a quantitative estimation of the amount of the intercalated alkaline metals was not performed, a certain amount of exfoliated layers and corresponding sodium and potassium sulfide phases were found within the shells of these nanoparticles. This finding raises the question, whether all hollow sulfide nanostructures are unstable under intercalation due to the expansion of interlayer spacing or due to the excess amount of alkaline metal atoms in the outermost parts of the nanoparticle. In order to study this problem we have considered the stability of multiwalled $MoS_2$ nanotubes upon Li intercalation into the van der Waals gap between the walls.

For multiwalled nanotubes atomistic quantum mechanical calculations to study the intercalation are computationally inaccessible. Therefore, we extend here a phenomenological model, that we previously developed and applied for the comparative study of the stability of nanotubes and nanostripes[51] or nanooctahedra and fullerene-like particles.[52]

In a first approximation the expansion or the compression of a cylindrical nanotube may be described by the change of its circumference or radius and a spring constant as in Hooke's law. Then, the excess in the strain energy $\Delta E_{str}$ for a single-walled nanotube, changing its radius may be written as:

$$\frac{\Delta E_{str}}{N} = \frac{Yw}{2\rho R_0^2}(\Delta R)^2, \quad (1)$$

where $Y$, $w$ and $\rho$ are the Young modulus, the thickness (the size of the van der Waals gap) and the number of atoms per surface area ($\rho = \sqrt{3}/6a^2$) for the corresponding flat monolayer. $R_0$ and $2\Delta R$ are the equilibrium radius of nanotube and the change of the radius upon expansion. $N$ is the number of atoms within the unit cell of the nanotube. Considering the case of a multiwalled nanotube with an odd number $k=2i+1$ of walls, we make next assumptions for simplicity:

i) Since the stress distribution between the walls within multiwalled nanotube remains unknown, the stress upon intercalation is viewed in a way, that the middle wall does not change its radius, while outer and inner walls increase and decrease their radii, respectively.

ii) The intercalation process for every nanotube causes a deformation with the uniform change of the radius along the circumference, which depends on $x$.

The total excess in the strain energy for intercalated a multiwalled nanotube relative to the pristine one can be obtained by:

$$\frac{\sum_{-i}^{i}\Delta E_{str}}{\sum_{-i}^{i}N}=\frac{Yw}{2\rho}\frac{\sum_{-i}^{i}\frac{(\Delta R)^2}{R_0}}{\sum_{-i}^{i}R_0},\quad (2).$$

As one can see from Fig. 5 the dependence of the change of the volume of a unit cell $\Delta V/V_0$ as a function of $x$ is nearly linear and the modulus of the corresponding change $\Delta V$ for the $i$-th shell:

$$\frac{\Delta V}{V_0}=a_x x\text{, and } |\Delta R|=a_x xwi \quad (3),$$

where $a_x$ is obtained to 0.2836 from our DFT calculations of the cell volumes for the bulk $Li_xMoS_2$ intercalates. It allows rewrite Eq. (2) in the form:

$$\frac{\sum_{-i}^{i}\Delta E_{str}}{\sum_{-i}^{i}N}=\frac{\sqrt{3}Yw^3}{a^2}a_x^2x^2\frac{\sum_{-i}^{i}\frac{i^2}{R_0}}{\sum_{-i}^{i}R_0}. \quad (4)$$

The values $Y$ = 238 GPa, $w$ = 6.15 Å and $a$ = 3.16 Å for 2H-$MoS_2$ can be available from experimental data.[35,53] The plots of the total excess of strain energy per atom $\Delta E_{str}$ for multiwalled $MoS_2$ nanotubes intercalated up to the $Li_{1.0}MoS_2$ composition are calculated using Eq. (4) and drawn in Fig. 6. Like the strain energy $E_{str}$ itself,[51] this function is inversely proportional to the $R^2$ and is proportional to the number of walls. It rapidly increases with decreasing radius $R$. However, synthesized $MoS_2$ nanotubes have radii in the order of $10^2$Å, and correspondingly the values for $\Delta E_{str}$ will not exceed a few meV/atom. Obviously, such small strain energy contributions cannot play an essential role in the destabilization of the tubular structure. These estimations show, that the walls of sulfide nanotubes are mechanically sufficiently resistant to the expansive deformation after Lithium intercalation. Therefore, the features related with morphological transformations of these nanoparticles, like exfoliation after interaction with alkali-metals, may be attributed only to the chemical nature of the bulk sulfide itself, particularly, to the ability for a chemical reduction.

**4. Conclusions**

We have studied in detail the Li intercalation into various Molybdenum disulfide allotropes. It could be demonstrated that there is still potential for improving the Li intercalation, and in this way to enlarge the capability of layered chalcogenides as electrode materials. The allotropic modifications of $MoS_2$ present a rich manifold for possible Li intercalation. For the first time a comparative analysis of the stability of allotropic 2H- and 1T-based $Li_xMoS_2$ intercalates was performed as a combined function of the coordination environment and arrangement of the intercalating atoms for a wide range of Li concentrations.

In contrast to Lithium Sulfur compounds, as for example $Li_2S$, where a tetrahedral coordination of the Li atoms is preferred, we recognize the octahedral coordination of Li atoms in the van der Waals gap as the most favorable one.

The Li content determines the most favorable type of $MoS_2$ lattice as hosting matrix. For $x < 0.4$ the $Li_xMoS_2$ intercalates with a trigonal prismatic coordination of Mo atoms is preferred. I.e., the hexagonal 2H-$MoS_2$ lattice is the most stable phase. At higher Li contents a monoclinic $Li_xMoS_2$ lattice with a distorted octahedral coordination of Mo atoms and a $2a \times 2a$ superstructure becomes more stable. The concentration dependence of the formation energies for the most stable phases indicate, that Li intercalation and deintercalation within the 2H-$MoS_2$ matrix can be performed smoothly in both directions, while deintercalation in the monoclinic $Li_{1.0}MoS_2$ structure is endothermic and requires ~0.8 eV/Li-atom to reversibly achieve a Li content of $x < 0.4$.

Many features of the relative stability of hexagonal 2H-, 1T- and monoclinic $Li_xMoS_2$ can be easily understood from the electronic structure of these phases, even within the concepts of crystal field theory. All $Li_xMoS_2$ allotropic forms have metallic-like character, excluding the “parent” phase, the Li-free 2H-$MoS_2$ and the monoclinic $Li_{1.0}MoS_2$ intercalate, which are the semiconductors.

Finally, we find that an increasing of the Li intercalation rate leads to the proportional increase of the volume of both the hexagonal 2H and the monoclinic $Li_xMoS_2$ intercalates, roughly in a similar way and mainly due to the expansion of the interlayer distance. For the most heavily intercalated $Li_{1.0}MoS_2$, the volume increase reaches ~15-20% compared with the corresponding parent $MoS_2$ allotrope. Our estimations evidence that even such expansions cannot promote an exfoliation of closed $MoS_2$ nanostructures. Lithium intercalation of $MoS_2$ nanostructures is limited only by the chemical stability of the sulfide against reduction.

**Acknowledgments.** We acknowledge financial support from the European Union within FP 7, Projects "MAHEATT", Contract. No. 227541, and INTIF 226639. A.E. thanks grant RFBR 11-03-00156-a.

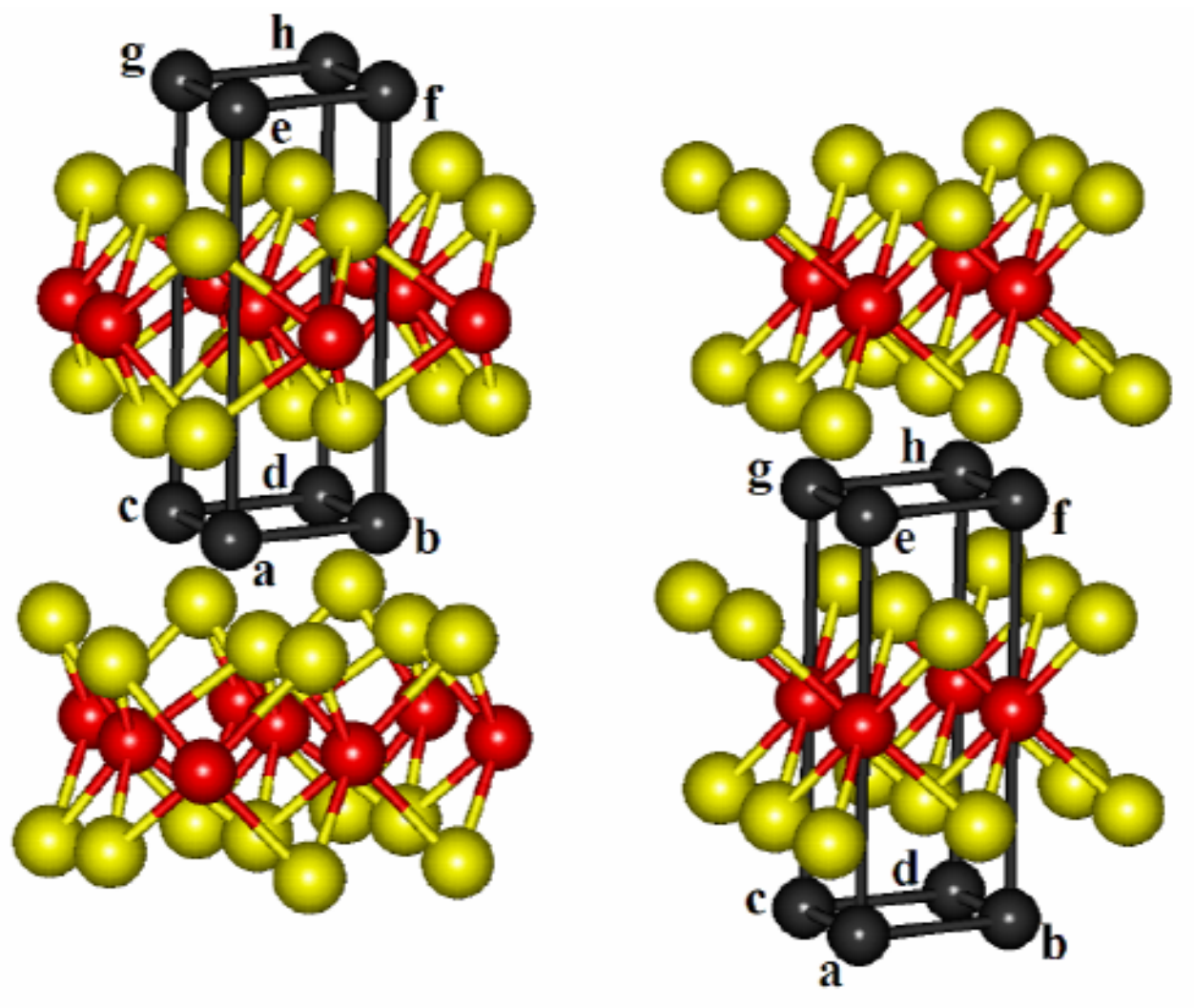

**Figure 1.** Fragments of the bulk structures for hexagonal 2H-$MoS_2$ (*on the left*) and hexagonal 1T-$MoS_2$ (*on the right*) composed of prismatic or octahedral units $MoS_6$, respectively. Labels for octahedral interstitials within the van der Waals gap of the 2x2x1 supercell of 2H-$MoS_2$ and the 2x2x2 supercell of 1T-$MoS_2$ are also shown.

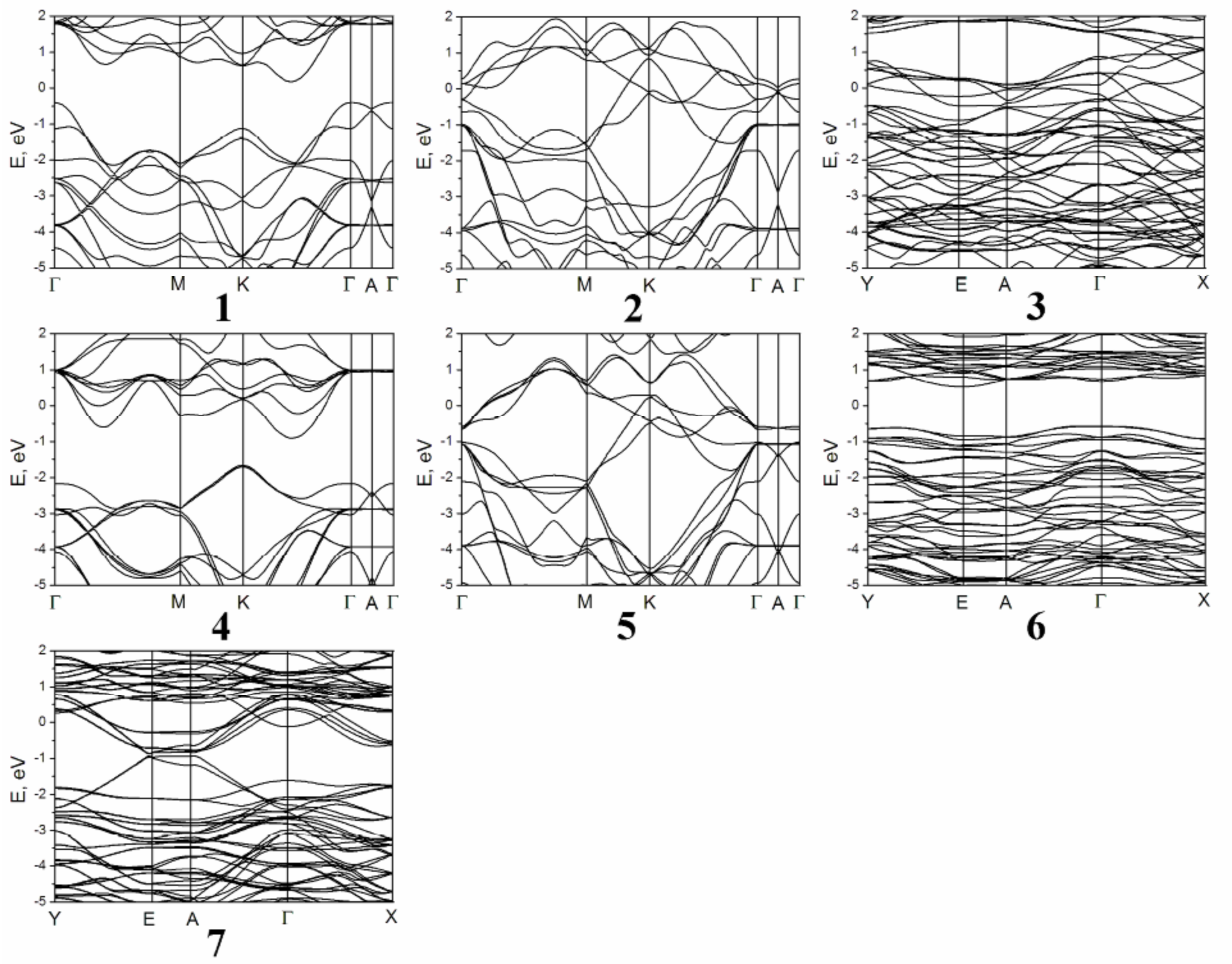

**Figure 2.** Band structures, calculated for 2H-$MoS_2$ (1), 1T-$MoS_2$ (2), monoclinic $MoS_2$ (3), 2H-$Li_{1.0}MoS_2$ (4), 1T-$Li_{1.0}MoS_2$ (5), monoclinic $Li_{1.0}MoS_2$ (6) and orthorhombic distorted 2H-$Li_{1.0}MoS_2$ (7). The Fermi level is set to 0.0 eV.

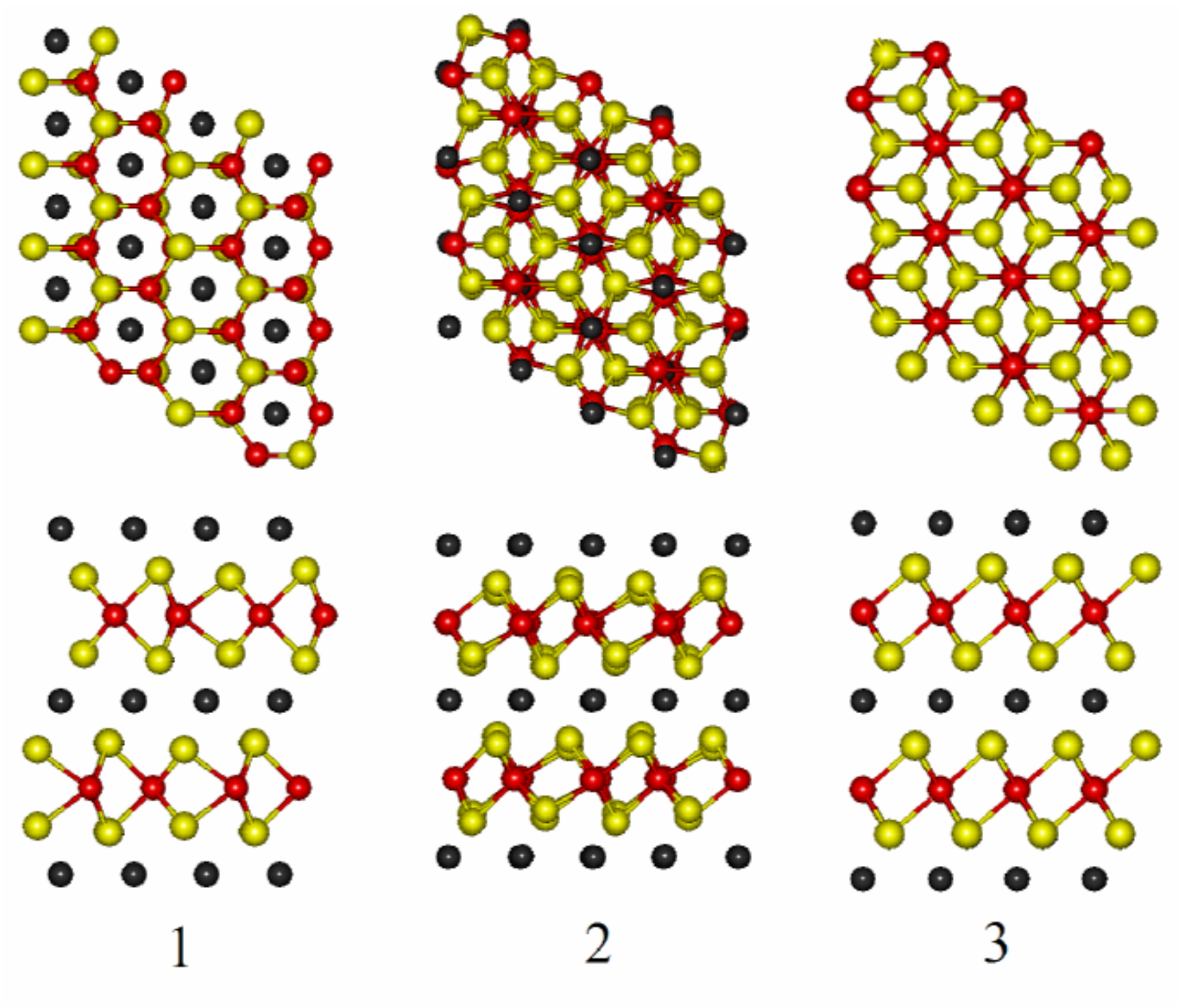

**Figure 3.** Fragments of optimized crystal structures for $Li_{1.0}MoS_2$ intercalates: stable orthorhombic compound with distorted sublattice of 2H-$MoS_2$, i.e. preserving the prismatic coordination of Mo atoms (*1*), stable monoclinic (*2*) and metastable hexagonal 1T structures (*3*) with octahedral coordination of Mo atoms. Top and side views are shown.

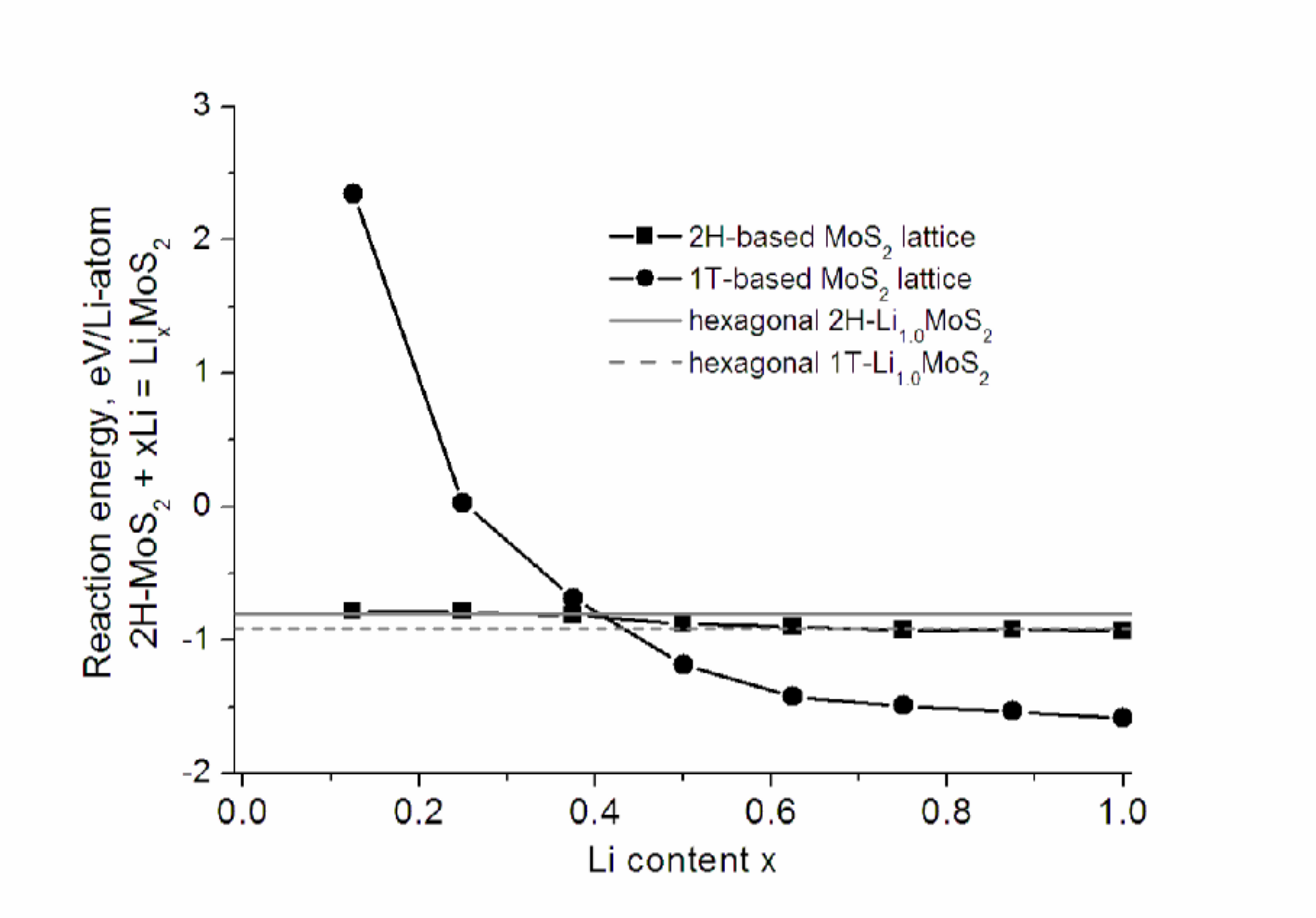

**Figure 4.** Calculated reaction energies of the formation of $Li_xMoS_2$ intercalates from the bulk of 2H-$MoS_2$ and bcc-Li as a function of the Lithium content *x* and the lattice structure.

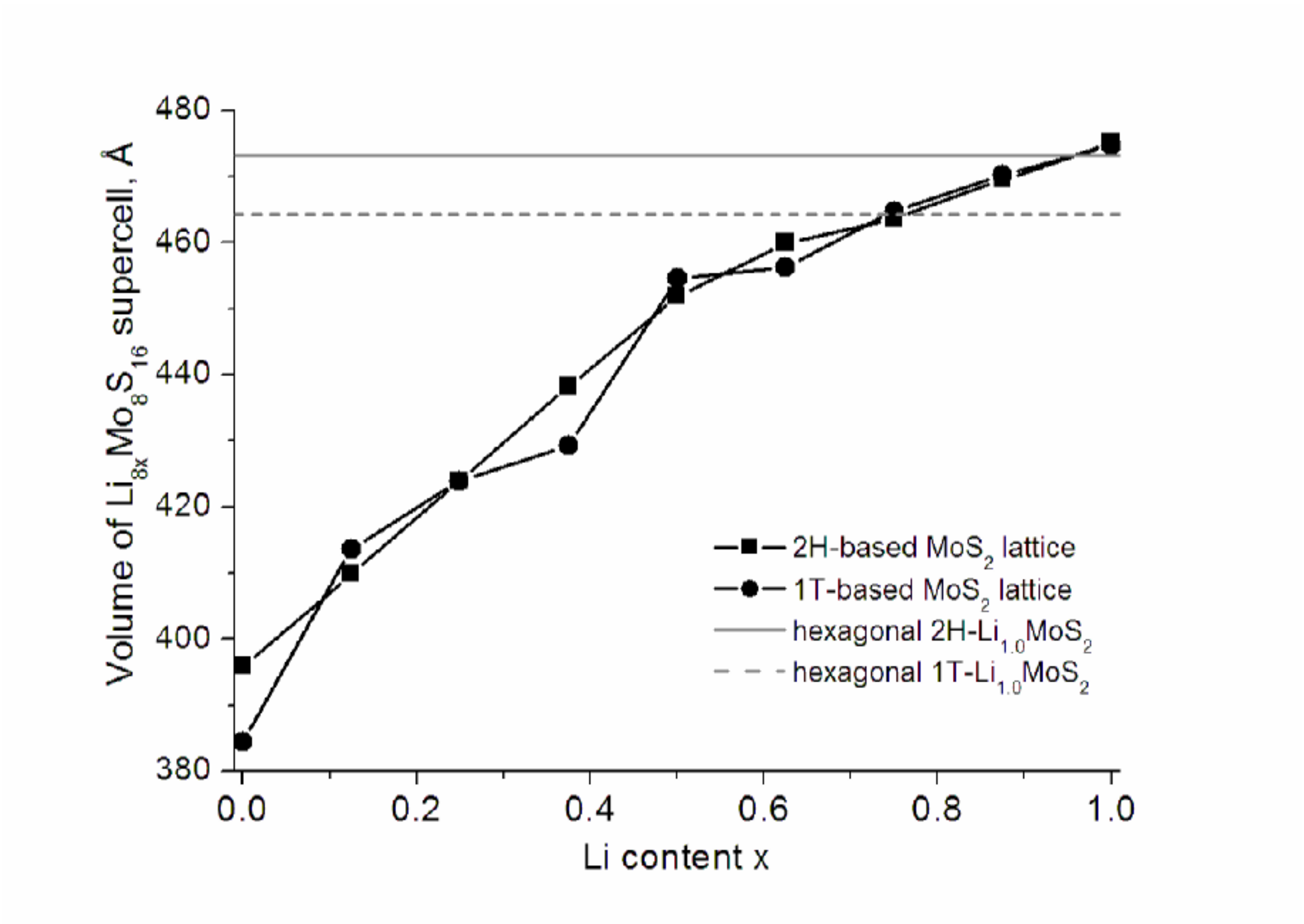

**Figure 5.** Calculated volume of a model supercell of $Li_xMoS_2$ intercalates as a function of the Lithium content x and the lattice structure.

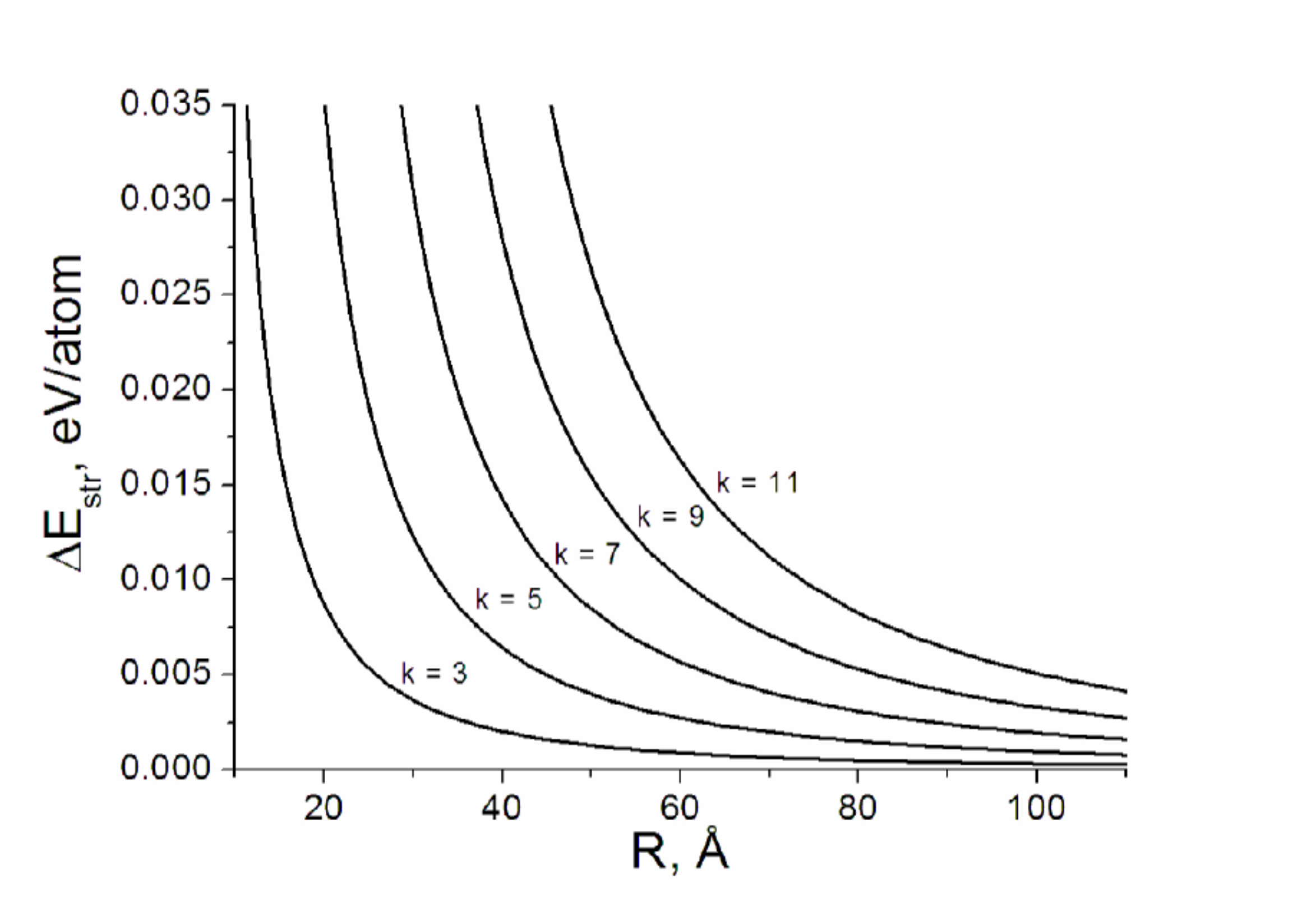

**Figure 6.** The difference in the strain energies between intercalated $Li_{1.0}MoS_2$ and pure $MoS_2$ *k*-walled nanotubes, estimated with the phenomenological model depending on the radius *R* of the middle wall.

Table 1. Calculated formation energies and lattice parameters for a set of $Li_xMoS_2$ intercalates with hexagonal symmetry of the hosting matrix (every unit cell contains two stoichiometric $MoS_2$ units)

| Composition | Intercalation site (t – tetrahedral, o - octahedral) | Lattice parameter, Å | | Formation energy, eV per Li-atom** |
|---|---|---|---|---|
| | | a | c | |
| 2H-$Li_{1.0}MoS_2$ | tt* | 3.1546 | 14.2025 | -0.7157 |
| 2H-$Li_{1.0}MoS_2$ | tt | 3.1562 | 14.2262 | -0.7083 |
| 2H-$Li_{1.0}MoS_2$ | oo | 3.1464 | 13.8007 | -0.8056 |
| 2H-$Li_{1.0}MoS_2$ | to | 3.1570 | 13.9836 | -0.7570 |
| 2H-$Li_{0.5}MoS_2$ | t | 3.1377 | 12.9464 | -0.6694 |
| 2H-$Li_{0.5}MoS_2$ | o | 3.1344 | 12.7128 | -0.7550 |
| 2T-$Li_{1.0}MoS_2$ | tt* | 3.1986 | 13.9835 | -0.5422 |
| 2T-$Li_{1.0}MoS_2$ | tt | 3.1970 | 13.9040 | -0.6485 |
| 2T-$Li_{1.0}MoS_2$ | oo | 3.2025 | 13.1331 | -0.8543 |
| 2T-$Li_{1.0}MoS_2$ | to | 3.2074 | 13.4493 | -0.7454 |
| 2T-$Li_{0.5}MoS_2$ | t | 3.1694 | 12.6710 | +0.4432 |
| 2T-$Li_{0.5}MoS_2$ | o | 3.1629 | 12.2351 | +0.0560 |
| 1T-$Li_{1.0}MoS_2$ | tt | 3.1980 | 14.0670 | -0.4684 |
| 1T-$Li_{1.0}MoS_2$ | oo | 3.2046 | 13.0073 | -0.9150 |
| 1T-$Li_{1.0}MoS_2$ | to | 3.2007 | 13.5514 | -0.6893 |
| 1T-$Li_{0.5}MoS_2$ | t | 3.1877 | 12.4713 | +0.3993 |
| 1T-$Li_{0.5}MoS_2$ | o | 3.1879 | 11.9990 | -0.0449 |

* atoms from both Li layers occupy tetrahedral interstitials near one and the same layer

** following the reaction equation 2H-$MoS_2$ + xLi → $Li_xMoS_2$